\documentclass[conference]{IEEEtran}
\IEEEoverridecommandlockouts
\usepackage{cite}
\usepackage{amsmath,amssymb,amsfonts}
\usepackage{algorithmic}
\usepackage{graphicx}
\usepackage{textcomp}
\usepackage{xcolor}
\usepackage{hyperref}
\def\BibTeX{{\rm B\kern-.05em{\sc i\kern-.025em b}\kern-.08em
    T\kern-.1667em\lower.7ex\hbox{E}\kern-.125emX}}
\begin{document}

\title{User-Centric AI Analytics for Chronic Health Conditions Management\\
}

\author{\IEEEauthorblockN{Aladdin Ayesh}
\IEEEauthorblockA{\textit{School of Natural and Computing Sciences} \\
\textit{University of Aberdeen}\\
Aberdeen, UK \\
aladdin.ayesh@abdn.ac.uk \\
\url{https://orcid.org/0000-0002-5883-6113}}

}

\maketitle

\begin{abstract}
The use of AI analytics in health informatics has seen a rapid growth in recent years. In this talk, we look at AI analytics use in managing chronic health conditions such as diabetes, obesity, etc. We focus on the challenges in managing these conditions especially with drug-free approaches due to the variations in individual circumstances. These variations directed the research into  user-centric approach leading to variety of research questions. In this short paper, we give examples from recent and current research work and conclude with what, in our opinion, to be the next steps and some remaining open research questions. 
\end{abstract}


\section{Introduction}

The use of AI analytics in health applications has seen a rapid growth in recent years. This is helped by advances in two areas. First, advances in sensor technologies including the wide spread use of wearable devices, e.g. smart watches, which enable continouse data gathering with minimum disruption to daily life. Second, advances in data analytic algorithms, e.g. deep learning, and high performance computers that allowed the processing of large datasets at speed that was not imaginable a decade ago. 

In this talk, we look at AI analytics use in managing chronic health conditions such as diabetes, chronic stress, dyslexia, etc. We focus on the challenges in managing these conditions especially with drug-free approaches due to the variations in individual circumstances. These variations directed the research into  user-centric approach leading to variety of research questions. In the following sections, we explore examples from our research work in this area. The talk concludes with what, in our opinion, to be the next steps and some remaining open research questions.

\section{Chronic Health Conditions}

Chronic health conditions can cover a wide range of diseases and disorders, which share common characteristics.  First, improvement in treatments and health care turned these diseases from terminal or seriously impairing to manageable conditions. Second, these conditions require active management plans to control their effects and to maintain quality of life for the sufferer. Thirdly, their impact and re-occurring manifestation may differ from person to person depending on several factors including personal and environmental. Finally, all of these conditions seem to have great psychological and emotional impact on their sufferers, which in turn may have an impact on the management of the condition and may lead to other health conditions of concern.

The common Examples of chronic conditions include diabetes, blood pressure, dyslexia, Alzheimer's, and cancer. Some of these are more widely researched than others in the context of technological solutions to managing them. We can view these conditions from variety of perspectives. Each perspective would emphasize certain characteristics of the a given health condition and potential approaches to control, treat, and manage the condition. The perspective from which we view the condition may also suggest the potential Artificial Intelligence (AI) technological solutions that can be employed to manage the condition and improve quality of life for the sufferer. 

\subsection{Nutrition Related}

There is a growing host of applications aimed towards managing nutrition related chronic health conditions, e.g. Diabetes Type 2. 
Diabetes as a health condition has been extensively researched \cite{Salvia2023,Ejiyi2023}. Since it is related to diet and weight, diet planning and weight management applications have seen exponential growth. However, many of these applications lack two key important aspects. First, there is limited, if any, nutritional research underpinning these applications limiting their scientific validation for long term effectiveness. Second, there is limited personalisation. The personalisation models are often preference models rather than a comprehensive user model that enables the understanding of the user's behavioural patterns and their impact on the user's  nutritional and health profile. A key challenge in this area is capturing the right amount and type of data from users and environment with minimum disruption to daily life. Techniques we have been exploring combine the use of gamification, wearable sensors, and self-reporting, in addition to nutritional dictionaries. The current work we are doing in this area is still at early stages and follows multiple streams. 


\subsection{Physiological Related}
Cancer \cite{Dragoni2023,Parimbelli2021} is an example of diseases that became chronic conditions due to the growing treatments available to control the disease. Unlike diabetes full recovery is possible but there is the risk of re-occurrence. This requires continuous monitoring. Developments in computer vision show promising potential for fast, accurate and less expensive diagnostic tools. 
Other physiological health conditions may require more user-centric approaches to fuse multi-modal data and provide explanations \cite{Zapala2021}.  

\subsection{Mental Related}
There is a growing recognition of the potential impact of AI technologies on mental health \cite{Lim2020,Wang2019,Ayesh2017} and well-being \cite{Schiff2020}. If this impact is well managed, it has a huge potential in addressing a large number of mental health issues and improving quality of life for millions on a daily basis. 


\section{Managing Chronic Conditions}
\begin{figure}[htbp]
		\includegraphics[width=\columnwidth,height=7cm]{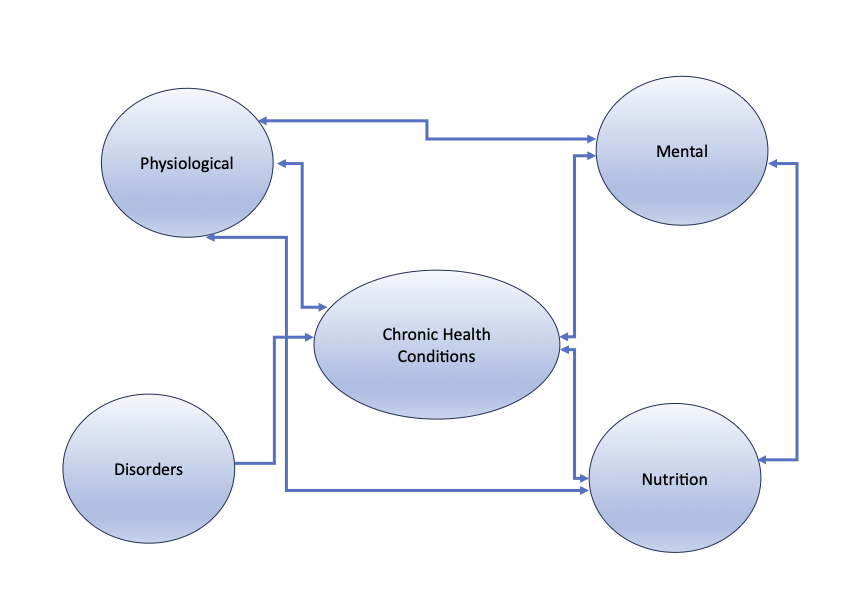}
	\caption{Chronic Health Conditions - Interlink View}
	\label{CHCinterlink}
\end{figure}

As stated in previous sections, chronic health conditions can be the result of physical, mental issues or combination of underlying causes.  The later is more often than none, and the consequences can be seen in many forms, e.g. a physiological condition will have a psychological impact similarly a mental condition is likely to manifest itself in physiological ailment.  Figure \ref{CHCinterlink} visualises the interlinked nature of the various potential causes of chronic health conditions. Thus, and as stated earlier, any categorisation of this kind is only helpful as far as emphasising a particular characteristics of the given condition in way of suggesting the requirements for specifying technological solutions. 

To progress our research in this area, we opted to focus on specific conditions within constraint settings. The first of these setting is educational systems. The second setting is in a more generalised form with potential applications in smart cities healthcare.


\subsection{In educational context}
Educational systems have been in use for years and became essential requirement during the COVID-19 pandemic. Their wide spread use gives us the opportunities to address variety of conditions and AI algorithms that can be generalised beyond the original purpose of complementing educational systems \cite{Ayesh2017,MarcoGimenez2016,ArevalilloHerraez2014}. For example emotional stress is a common condition impact technology users. Some acute forms can be observed during the use of educational systems \cite{Lim2017,Lim2020} which enabled us to provide adaptive mechanisms by which the system responds to the user needs and reduce the temporary stress. Our approach combined quantitative and qualitative data analytics informing a rule-based system to perform the adaptation. 

Dyslexia is another condition that often impact student performance both in class rooms and while using online educational systems \cite{Wang2019,Wang2017}.  For this condition, a more sophisticated optical sensor system, i.e. eye-gazing goggles, were necessary to assess student's motivation while using online educational systems. The eye tracking data was combined with other sensory readings, namely EEG \cite{Wang2022}, to develop a framework that enables personalised services in the educational context. In section \ref*{PAIM} we revisit personalisation of AI models in more generalised form.



\subsection{Ageing population and smart cities context}
Another context for chronic health conditions management is assistive technologies that are becoming ever more necessary with ageing populations and the emergent of smart cities. Two relevant areas, which are currently focal topics are empathetic technologies and neural interfaces. There are multiple IEEE standards working groups developing standards and recommended practices in these two areas, e.g. \cite{Schiff2020} and \cite{Zapala2021}.



\section{Personalising AI Models}
\label{PAIM}
\begin{figure}[htbp]
		\includegraphics[width=\columnwidth,height=8cm]{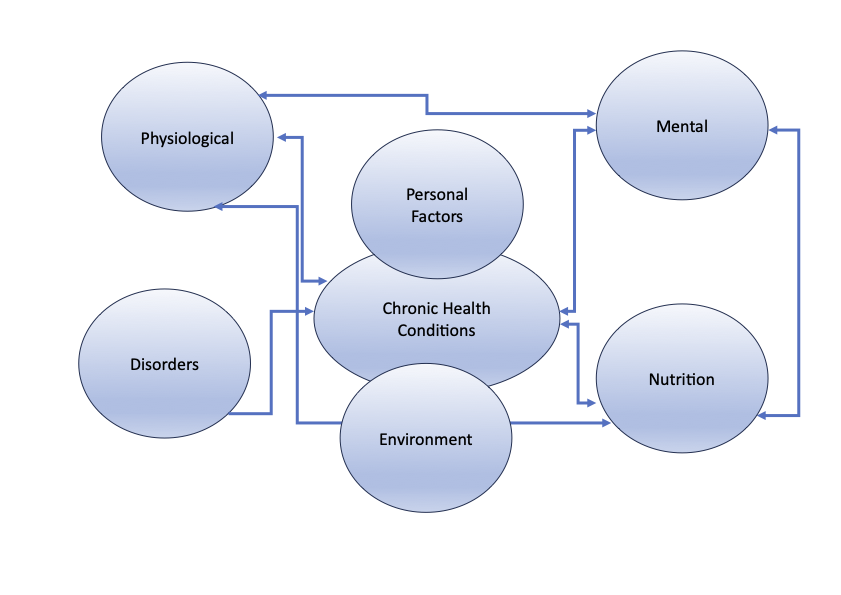}
	\caption{Chronic Health Conditions - Individual Centric Factors}
	\label{CHCPF}
\end{figure}

A key approach that underlines and motivates many AI advances, is to enable individualised or situated responses and interactions between individual user and AI enabled system. We adopted this as a principle in exploring and developing personalised AI models \cite{Ayesh2017,Ayesh2016,Ayesh2015,Ayesh2015a,Ayesh2014}. The aim is that when machine learning and analytic algorithms are used, the learned models would take into consideration more personal factors thus produce more relatable results from a user perspective than focusing on abstract metrics. The need for such personalised AI models becomes more evident in the context of chronic health conditions. Figure \ref{CHCPF} shows two sets of factors impacting individual cases within this context. 

This approach opens a wide range of questions about metrics and evaluation methods in machine learning and analytic algorithms similar to some already in existence where cognitive systems are concerned but extends beyond. 

\section{What is next?}
The next step is to develop an integrated  framework for mental and physical health. This framework could combine the various sensors already used in existing wearable devices to provide the building blocks for intelligent health monitoring systems for chronic health conditions management and support to deliver on the high expectations of AI positive impact in delivering health care \cite{Koutsouleris2022,Aerts2021}.

The need for better metrics and evaluation methods in evaluating machine learning and analytic algorithms becomes an urgent issue for research to address. It is not sufficient to evaluate the algorithms using abstract metrics such as recall and accuracy but instead there is a need for interpretation and explainable evaluation.  Thus, there is a need for practical translation of explainable and responsible AI to be embedded in practical applications to provide a better oversight of algorithmic AI, which continues to prove its invaluable practical uses with equally increasing concerns.

\section{Conclusion}
In this short paper, we have presented the personalisation requirements and potential benefits of AI technologies in managing chronic health conditions. By doing so, we have also presented the case for more personalised AI models to take into consideration the fine differences in personal and environmental factors that could impact greatly the chronic condition manifestation and health impact on a given individual.


\bibliographystyle{ieeetr}
\bibliography{chronichealth}

\end{document}